\RequirePackage{fix-cm}
\documentclass{svjour3}                     
\smartqed  
\usepackage{graphicx}
\begin{document}

\title{General Relativistic Razor-Thin Disks with Magnetically Polarized Matter
}


\author{Anamar\'ia Navarro-Noguera         \and
        F. D. Lora-Clavijo \and
        Guillermo A. Gonz\'alez 
}


\institute{Anamar\'ia Navarro-Noguera \at
              Grupo de Investigaci\'on en Relatividad y Gravitaci\'on, 
 Escuela de F\'isica, Universidad Industrial de Santander, A. A. 678,
Bucaramanga 680002, Colombia. \\
              \email{ana.navarro1@correo.uis.edu.co}           
           \and
            F. D. Lora-Clavijo \at
              Grupo de Investigaci\'on en Relatividad y Gravitaci\'on, 
Escuela de F\'isica, Universidad Industrial de Santander, A. A. 678,
Bucaramanga 680002, Colombia. \\
              \email{fadulora@uis.edu.co}
           \and
           Guillermo A. Gonz\'alez \at
              Grupo de Investigaci\'on en Relatividad y Gravitaci\'on, 
Escuela de F\'isica, Universidad Industrial de Santander, A. A. 678,
Bucaramanga 680002, Colombia. \\
              \email{guillermo.gonzalez@saber.uis.edu.co}
}

\date{Received: date / Accepted: date}

\maketitle

\begin{abstract}
The origin of magnetic fields in the universe still remains unknown and constitutes one of the most intriguing questions in astronomy and astrophysics. Their significance is enormous since they have a strong influence on many astrophysical phenomena. In regards of this motivation, theoretical models of galactic disks with sources of magnetic field may contribute to understand the physics behind them. Inspired by this, we present a new family of analytical models for thin disks composed by magnetized material. The solutions are axially symmetric, conformastatic and are obtained by solving the Einstein-Maxwell Field Equations for continuum media without the test field approximation, and assuming that the sources are razor-thin disk of magnetically polarized matter.  We find analytical expressions for the surface energy density, the pressure, the polarization vector, the electromagnetic fields, the mass and the rotational velocity for circular orbits, for two particular solutions. In each case, the energy-momentum tensor agrees with the energy conditions and also the convergence of the mass for all the solutions is proved.  Since the solutions are well-behaved, they may be used to model astrophysical thin disks, and also may contribute as initial data in numerical simulations. In addition, the process to obtain the solutions is described in detail, which may be used as a guide to find solutions with magnetized material in General Relativity. 
\end{abstract}

\section{Introduction}
\label{intro}

Magnetic fields have been measured in practically all celestial objects. For instance, they have been detected with significant strength in the Earth, stars, pulsars, the Milky Way, nearby galaxies, more distant (radio) galaxies, quasars and intergalactic space in clusters of galaxies \cite{2016cmf..book.....K}. However, the origin of the first magnetic fields in the Universe still remains unclear. Therefore, the standard process  known as $\alpha \omega$-dynamo, which assume that the magnetic fields in spiral galaxies come from the combined action of differential rotation and helical turbulence, does not explain the existence of magnetic fields in elliptical galaxies and clusters \cite{2002RvMP...74..775W}.  
The first suggestions about the presence of magnetic field in nearby galaxies were made in \cite{1958ApJ...128....9H}. In this work the authors,  based on the observations of twenty-one globular clusters in the Andromeda galaxy, M31, have found that interstellar polarization is presented. Moreover, in \cite{1967PASP...79..102A} the mean direction of the interstellar magnetic field  in the vicinity of the Sun was obtained more accurately. 
Magnetic fields in nearby galaxies have been detected by using different types of probes \cite{2013IAUS..294..213H}. The most classic one corresponds to the optical polarization, which has been used in recent observations to get a first detailed map of  the magnetic field strength for a quiescent molecular cloud, using the polarized light of stars behind the clouds \cite{2012ApJ...755..130M}.  A new method for measuring the large-scale structure of the Galactic magnetic field was presented in \cite{2014AJ....148...49P}, showing new results associated with its morphology.  On the other hand, observations of polarized emission of clouds and dust grains, in order to study inflationary cosmology and the Milky Way Galaxy's composition and magnetic field structure, have been achieved in the Galactic plane \cite{2011ApJ...741...81B}. 

In order to provide a theoretical model to explain magnetic fields in galaxies, many exact solutions of Einstein-Maxwell equations for thin disks with magnetic fields have been obtained.  For instance,  in \cite{LetelierExactRelatDisksMagneticFields} were found four exact solutions of hot disks; later in \cite{GarciaReyesGonzalez}, the counterrotating model for electrovacuum axially symmetric relativistic thin disks without radial stress was presented. Moreover in 
\cite{GarciaReyesGonzalez2009Int.J.Mod.PhysD18}, a family of counterrotating and rotating relativistic thin disks, based on a charged and magnetized Kerr-NUT metric, was constructed. Besides,  in \cite{GutierrezGonzalezInfiniteFamilyMorganMorgan} were obtained solutions of magnetized Morgan-Morgan Relativistic thin disks. In \cite{BallenGonzalez}, solutions of relativistic dust thin disks with halo and magnetic field were described. Later on, in 
\cite{2014-guttierrez-garcia-gonzalez} solutions of thin disks around black holes, under the influence of magnetic fields,  were found. In \cite{García-Reyes2014}, using the  magnetized Reissner-Nordstrom metric, thin disk solutions composed by perfect fluid and magnetic fields were constructed. Recently in \cite{2015GReGr..47...54G}, solutions for a conformastationary metric with magnetized disk-haloes sources and  solutions of rotating disks in magnetized haloes \cite{2015arXiv150401138G}, were analyzed. 

In the previously described works, the sources of the electromagnetic fields are charges and current densities. Magnetic fields generated by magnetically polarized material have not been studied enough. As far as we know, the only intents to include this kind of material were presented in \cite{monopolos,dipolosenschwarchild}, where the authors analyzed the electromagnetic fields originated by electric and magnetic dipole layers in a Schwarzschild spacetime. In these works the test field approximation is used, so the solutions are obtained by solving the Maxwell equations and overlapping the solution with the fixed background metric. Therefore, we consider of great importance to obtain exact solutions that correspond to relativistic thin disks with magnetic polarized source, due to the observational evidence of the magnetic dipoles and to contribute to the general study of relativistic thin disks.

Accordingly with the above considerations, in this work we describe how to obtain an infinite family of relativistic static thin disks composed by magnetically polarized matter, which are obtained by solving the Einstein-Maxwell equations for continuum media, without considering the test field approximation and assuming that the sources are razor thin disk of magnetically polarized matter. In order to obtain the solutions, we adopt the distributional approach for tensors, which has been widely used to study this kind of systems, see for instance \cite{2008PhRvD..78f4058G,2010PhRvD..82h4005L,2013PhRvD..87d4010G}. 
This paper is organized as follows: in section \ref{Sec:Ecuaciones}, we obtain the Einstein-Maxwell equations for a disk of electrically and magnetically polarized material. In section \ref{Sec:magnetizedDisk}, we assume a particular metric and an electromagnetic potential of a magnetized material, the equations are solved and we find analytical expressions for an infinite family of solutions. In section \ref{Sec:Results}, we analyze the behavior of the  density, pressure, magnetization vector, magnetic fields,  the velocity curves for circular orbits for the first two members of the family, and also the convergence of the mass of all the models.

\section{Einstein-Maxwell Equations for an electromagnetically polarized disk}
\label{Sec:Ecuaciones}

The Einstein-Maxwell equations for a continuum media, in Heaviside-Lorentz geometrized units such that $c= 8\pi G = \mu_0 = \epsilon = 1$, are 
\begin{eqnarray}
 & & G_{\alpha \beta}  =  T^M_{\alpha \beta} + T^F_{\alpha \beta} +  T^{FM}_{\alpha \beta} \, , \label{eq_einstein}  \\
 & & {F^{\alpha \beta}}_{;\beta} = {M^{\alpha \beta}}_{;\beta} \, , \label{eq_maxwell}
\end{eqnarray}
where the energy-momentum tensor has three components, the first one corresponds to the mass contribution $T^ {M}_{\alpha \beta}$, the second one is due to the electromagnetic fields, $T^F_{\alpha \beta}$, and the third one represents the electromagnetic interaction with the polarized matter $T_{\alpha \beta}^{FM}$ \cite{Grot_JMathPhys-11-109_1970,1978JMP....19.1198M},
\begin{eqnarray}
& & T_{\alpha \beta}^F = F_{\alpha \mu} {F_{\beta}}^{\mu} - \frac{1}{4}g_{\alpha \beta} F_{\mu \nu} F^{\mu \nu} \, , \label{TF} \\
& & T^{FM}_{\alpha \beta} = -F_{\alpha \mu} {M_{\beta}}^{\mu} \, , \label{TFM}
\end{eqnarray} 
where $F_{\alpha \beta}$ is the electromagnetic tensor and $M_{\alpha \beta}$ is the polarization-magnetization tensor, respectively given by
\begin{eqnarray}
 & & F_{\alpha \beta} = A_{\beta, \alpha} - A_{\alpha, \beta} \, , \label{F=A-A}  \\
 & & M^{\alpha \beta} = P^{\alpha} u^{\beta} - P{^\beta} u^{\alpha} + \epsilon^{\alpha \beta \mu \nu}M_{\mu}u_{\nu} \, ,
\end{eqnarray} 
with $A_\alpha$ the electromagnetic potential, $P_{\alpha}$ the electric polarization, $M_{\alpha}$  the magnetic polarization, $u_{\alpha}$ the 4-velocity of the observer co-movil with the fluid, $\epsilon^{\alpha \beta \mu \nu}$ is the Levi-Civita tensor and the comma stands for the usual derivative. 

In order to obtain a disk-like solution, we only consider mass located in the surface $z=0$, and we assume that the metric $g_{\alpha \beta}$ has symmetry reflection through the disk surface
\begin{eqnarray}
g_{\alpha \beta}(r,z) = g_{\alpha \beta}(r,-z) \, ,
\end{eqnarray}
in such a way that for $z\neq 0$, 
\begin{eqnarray}
{g_{\alpha \beta}}_{,z}(r,z) = -{g_{\alpha \beta}}_{,z}(r,-z) \, . \label{eq_g_refsim}
\end{eqnarray}
In addition, we assume that the metric tensor is continuous in the region of the disk, which can be written through the jump function
\begin{eqnarray}
\left[ g_{\alpha \beta}  \right] = \left. g_{\alpha \beta} \right|_{z=0^+} - \left. g_{\alpha \beta} \right|_{z=0^-} = 0 \, ,
\end{eqnarray}
where the upper-index $+$ and $-$ stand for the space above and below the disk surface. The gravitational field of the disk is modeled by a discontinuity in the first derivative of the metric tensor through the disk given by
\begin{equation}
b_{\alpha \beta} = [g_{\alpha \beta,z}] = {g_{\alpha \beta,z}}|_{z=0^+} -  {g_{\alpha \beta,z}}|_{z=0^-}  = 2g_{\alpha \beta , z}|_{z=0^+} \, , \label{Discontg}
\end{equation}
where the reflection symmetry (\ref{eq_g_refsim}) has been taken into account. 

Moreover, we assume that the electromagnetic potential has reflexion symmetry through the disk surface
\begin{equation}
A_{\alpha}(r,z) = A_{\alpha}(r,-z) \, , 
\end{equation}
in such a way that for $z \neq 0$,
\begin{equation}
A_{\alpha ,z}(r,z) = -A_{\alpha ,z}(r,-z) \, .
\end{equation}
Additionally, this potential is continuous through the region of the disk, so that its jump is zero
\begin{equation}
[A_{\alpha }] = {A_{\alpha}}|_{z=0^+}  - {A_{\alpha}}|_{z=0^-} = 0 \, .
\end{equation}
It is worth mentioning that $A_{\alpha }$ has a finite discontinuity in its first normal derivative, which can be represented as
\begin{equation}
[A_{\alpha, z}] = {A_{\alpha,z}}|_{z=0^+}  - {A_{\alpha,z}}|_{z=0^-}= 2 {A_\alpha}|_{z=0^+} , \label{DiscontA}
\end{equation}
where the reflection symmetry has been taken into account. 

Furthermore, in order to use the distributional approach for tensors 
\cite{Lichnerowicz1971,Papetrou_Surfacelayersofmatter1968,Taubspacetimeswithdistribution}, the metric $g_{\alpha \beta}$ can be expressed as
\begin{eqnarray}
g_{\alpha \beta} = g_{\alpha \beta}^+\Theta(z) + g_{\alpha \beta}^-\left\lbrace 1 - \Theta(z) \right\rbrace \, ,  \label{eq_g_D}
\end{eqnarray}
 where the Heaviside distribution is defined using the half-maximum convention  
\begin{equation}
\Theta(z) = \left\lbrace \begin{array}{cc}  1  & \ \ \textrm{for} \ \ z>0, \\ \frac{1}{2} & \ \ \textrm{for} \ \ z= 0,  \\ 0 & \ \ \textrm{for} \ \ z<0. \end{array} \right. 
\end{equation}
Following the same approach, the Einstein tensor can be divided into the three regions of the space as
\begin{eqnarray}
G_{\alpha \beta} =   G_{\alpha \beta}^+\Theta(z) + G_{\alpha \beta}^-\left\lbrace 1 - \Theta(z) \right\rbrace + Q_{\alpha \beta} \delta(z) \, , \label{G}   
\end{eqnarray}
where in this case, $Q_{\alpha \beta}$ stands for the Einstein tensor on the disk, and $\delta(z)$ is the Dirac distribution. Accordingly, the Ricci tensor can be expressed as
\begin{equation}
R_{\alpha \beta} = R_{\alpha \beta}^+ \Theta(z) + R_{\alpha \beta}^- \left\lbrace 1 - \Theta(z) \right\rbrace + H_{\alpha \beta} \delta(z) , \label{Ricci}
\end{equation}
where $H_{\alpha \beta}$ is the Ricci tensor on the disk. 
In regards to this discretization and the definition of the Einstein tensor, $G_{\alpha \beta}^\pm$ and $Q_{\alpha \beta}$ can be written as 
\begin{eqnarray}
G_{\alpha \beta}^\pm  =  R_{\alpha \beta}^\pm - \frac{1}{2} g_{\alpha \beta}R^\pm \, , \\
Q_{\alpha \beta} =  H_{\alpha \beta} - \frac{1}{2}g_{\alpha \beta}H \, ,  
\end{eqnarray}
being $H = g^{\alpha \beta} H_{\alpha \beta}$ the Ricci scalar on the disk with $H_{\alpha \beta}$ given by
\begin{eqnarray} 
H_{\alpha \beta} = \frac{1}{2}\left( b^z_{\ \alpha} \delta^z_{\ \beta} + b^z_{\ \beta} \delta^z_{\ \alpha} - b^\mu_{\ \mu} \delta^z_{\ \alpha} \delta^z_{ \ \beta} - g^{zz}b_{\alpha \beta}  \right)  \, . 
 \end{eqnarray}
This last expression was derived according to the discretization of the metric tensor (\ref{eq_g_D}) and all quantities are evaluated at $z=0$ (see \cite{Taubspacetimeswithdistribution}).  

On the other hand, the distributional approach for the components of the energy-momentum tensor $T_F^{\alpha \beta}$, and the components of the polarization tensor  are given as 
\begin{eqnarray}  
\label{TFD} T_F^{\alpha \beta} &=& {T_F^{\alpha \beta}}^+ \Theta(z) + {T_F^{\alpha \beta}}^-\left\lbrace 1 - \Theta(z)  \right\rbrace \, ,\\      
M^{\alpha \beta} &=& \Pi^{\alpha \beta} \delta(z) \, , \label{MD}
\end{eqnarray}
where $\Pi^{\alpha \beta}$ is the polarization tensor on the disk, since the material is placed only on the plane $z=0$. In consequence, the electric and magnetic polarization vectors of the disk can be written as
\begin{eqnarray}
  P_\alpha &=& u^\beta \Pi_{\beta \alpha}  \, , \qquad M^\alpha = \frac{1}{2} \varepsilon^{\alpha \beta \mu \nu} \Pi_{\mu \nu} u_\beta \, , \label{VecM}
\end{eqnarray}
where $u_\beta $ is the 4-velocity of the observer and $\varepsilon$ is the Levi-Civita tensor.    
  Now, by replacing  (\ref{TFD}) and (\ref{MD}) in equation (\ref{TFM}), we obtain an expression for the components of the energy momentum tensor $T^{\alpha \beta}_{FM}$, due to the electromagnetic interaction, as follows 
\begin{eqnarray}
T^{\alpha \beta}_{FM} = \tau_{FM}^{\alpha \beta} \delta(z) \, , \label{TFMD} 
\end{eqnarray}
with
\begin{equation}
\tau_{FM}^{\alpha \beta}  = {\bar{F}^\alpha}_{\ \mu} \Pi^{\mu \beta} ,
\end{equation}
and
\begin{equation}
{\bar{F}^\alpha}_{\ \mu} = \frac{ {F^{\alpha} _{ \ \mu}}^+ + {F^{\alpha}_{\ \mu}}^-  }{2} \, , 
\end{equation}
where all quantities are evaluated in $z=0$. The component due to matter would the be 
\begin{eqnarray}
T_M^{\alpha \beta} =  \tau^{\alpha \beta}_M \delta(z) \, , \label{TMD} 
\end{eqnarray}
where $\tau^{\alpha \beta}_M$ represents the matter component tensor on the disk, since all matter is located there. 

In order to write the Maxwell's Equations (\ref{eq_maxwell}), using the distributional approach, we first write the covariant derivative of the tensors $F^{\alpha \beta}$ and $M^{\alpha \beta}$, as follows
\begin{eqnarray} 
{\sqrt{-g}F^{\alpha \beta}}_{; \beta} &=& \left(\sqrt{-g}F^{\alpha \beta}\right)_{, \beta}  = {\hat{F}^{\alpha \beta}}_{\ \ \ , \beta}  \, , \label{eq_F_cov} \\
{\sqrt{-g}M^{\alpha \beta}}_{; \beta} &=& \left(\sqrt{-g}M^{\alpha \beta}\right)_{, \beta} = {\hat{M}^{\alpha \beta}}_{\ \ \ , \beta}   \, , \label{eq_M_cov} 
\end{eqnarray}
where $g=\textrm{det}(g_{\alpha \beta})$, $\hat{F}^{\alpha \beta}=\sqrt{-g}F^{\alpha \beta}$ and $\hat{M}^{\alpha \beta}=\sqrt{-g}M^{\alpha \beta}$. Moreover, by using the distributional approach the discretization for the derivative $ {\hat{F}^{\alpha \beta}}_{\ \ \ , \beta} $ takes the form
\begin{eqnarray}
{\hat{F}^{\alpha \beta}}_{\ \ \ , \beta} &=& {\hat{F}^{\alpha \beta \ + }_{\ \ \ ,\beta}} +  {\hat{F}^{\alpha \beta \ - }_{\ \ \ ,\beta}}  \left\lbrace 1 - \Theta(z) \right\rbrace   + \left[\hat{F}^{\alpha \beta}\right] \delta ^z_{\ \beta} \delta(z), \label{FhatD} 
\end{eqnarray}
in such a way that equation (\ref{eq_F_cov}) can be written as follows 
\begin{eqnarray}
\sqrt{-g} {F^{\alpha \beta}}_{; \beta} &=&\left( \hat{F}^{\alpha \beta}_{\ \ \ ,\beta} \right)^{\pm} + \left[\hat{F}^{\alpha \beta}\right] \delta ^z_{\ \beta} \delta(z), \label{FtD} 
\end{eqnarray}
On the other hand, $\hat{M}^{\alpha \beta}_{\ \ \ ,\beta}$ is obtained from the derivative of equation (\ref{MD}) 
\begin{eqnarray}
{\hat{M}^{\alpha \beta}}_{\ \ \ ,\beta} &=& \hat{\Pi}^{\alpha \beta}_{\ \ \ ,\beta } \delta(z) + \hat{\Pi}^{\alpha z} \delta '(z), 
\end{eqnarray}
so we can write equation (\ref{eq_M_cov}) as
\begin{eqnarray}
\sqrt{-g} {M^{\alpha \beta}}_{; \beta} &=&  \hat{\Pi}^{\alpha \beta}_{\ \ \ ,\beta } \delta(z) + \hat{\Pi}^{\alpha z} \delta '(z), \label{MtD}
\end{eqnarray} 
where $\hat{\Pi}^{\alpha \beta}=\sqrt{-g}\Pi^{\alpha \beta}$. Then, substituting (\ref{FtD}), (\ref{MtD}), (\ref{TMD}), (\ref{TFMD}) and (\ref{TFD}) into  (\ref{eq_einstein}) and (\ref{eq_maxwell}), the electrovacuum field equations are given by
\begin{eqnarray}
 G_{\alpha \beta}^\pm = \left( T_{\alpha \beta}^F\right)^\pm \, , \qquad 
(\sqrt{-g} \ F^{\alpha \beta})_{, \beta}^\pm  = 0 \, , \label{eq: F^D} 
\end{eqnarray} 
for $z \ge 0$ and $z \le 0$ and 
\begin{eqnarray}
& & {Q_{\alpha \beta}} = \tau_{\alpha \beta}^M + {\tau_{\alpha \beta}^{FM}} \, , \label{eq: Einstein disco}  \\
& &  \left[ \sqrt{-g} \ F^{\alpha z} \right] =  (\sqrt{-g} \ \Pi^{\alpha \beta})_{,\beta} \, ,  \label{eq: [F]}  \\
& & \sqrt{-g} \ F^{\alpha z} = 0 \, ,  \label{hat pi}  
\end{eqnarray} 
for $z=0$.

The physical characteristics of the solution are described by the energy density and pressures. Therefore is necessary to obtain the surface energy-momentum tensor by making use of equation (\ref{eq: Einstein disco})
\begin{eqnarray}
S_{\alpha \beta} = \int T^M_{\alpha \beta} \mathrm{d}s_n = \sqrt{g_{zz}} \left( Q_{\alpha \beta} -  \tau_{\alpha \beta}^{FM}  \right) \label{Tsup} \, ,
\end{eqnarray}
where $ds_n = \sqrt{g_{zz}}dz$ is the arc length in $z$-direction. Furthermore, those physical quantities need to be analyzed in a co-movil observer. Therefore, we introduce the orthonormal tetrad of the ``Locally Static Observer" \cite{KatzBicakLynden_Disksourcesforconformastationarymetrics}, defined by the relations 
\begin{eqnarray}
& & e^\alpha_{(t)} =  e^{-\psi} \delta^\alpha_t \, , \quad
e^\alpha_{(r)} =  e^{\psi} \delta^\alpha_r \, ,  \quad \nonumber \\
& & e^\alpha_{(\phi)} =  e^{\psi} \delta^\alpha_\phi /r \, , \quad 
e^\alpha_{(z)} =  e^{\psi} \delta^\alpha_z \, , \label{tetrada}
\end{eqnarray}
 with $e^\alpha_{(t)}$ the 4-velocity of the observer $u{^\alpha}$, the surface energy-momentum tensor can be written as
\begin{eqnarray}
\hspace{-0.6cm} S_{\alpha \beta} = \sigma e_{\alpha}^{(t)} e_{\beta}^{(t)} +  p_r e_{\alpha}^{(r)} e_{\beta}^{(r)} + p_\phi e_{\alpha}^{(\phi)} e_{\beta}^{(\phi)} + p_z e_{\alpha}^{(z)} e_{\beta}^{(z)}, \label{Sdescomp}
 \end{eqnarray}
where $\sigma$ is the surface energy density, $p_r$ is the radial pressure, $p_\phi$ is the azimuthal pressure and $p_z$ is the normal pressure of the disk \cite{RelativistikToolkit}. 

\section{Magnetized disks in a conformastatic spacetime}
\label{Sec:magnetizedDisk}

In orden to obtain a solution to model a magnetized galactic disk, we choose the electromagnetic four potential as $A_\alpha = \left[0,0,A(r,z),0\right] $ and, for the sake of simplicity, the line element for a conformastatic spacetime as given by
\begin{eqnarray}
 ds^2 = -e^{2\psi} dt^2 + e^{-2\psi} \left( dr^2+r^2d\phi^2 + dz^2 \right) \, , \label{ds2}
\end{eqnarray}
where $\psi=\psi(r,z)$. On the other hand, equation (\ref{hat pi}) implies that the only non-zero components of the polarization-magnetization tensor are $\Pi_{r \phi}$ and $-\Pi_{\phi r}$, and with (\ref{VecM}), it can be seen that the only non-zero component of the magnetization vector is its normal component, which is written as $ M_z =  e^{2\psi} \Pi_{r \phi} /r $, such that, the asymmetry of the polarization-magnetization tensor implies 
\begin{eqnarray}
\Pi_{r \phi} = -\Pi_{\phi r} = e^{-2\psi}r M_z \, .
\end{eqnarray}

Moreover, the Einstein-Maxwell equations outside the disk (\ref{eq: F^D}) yield to the following system of equations 
\begin{eqnarray}
2\psi_{,r} \psi_{,z} r^2 + e^{2\psi} A_{,r} A_{,z } &=&  0 \label{eqs1 m} \, , \\ 
 2 \psi_{,r}^2 r^2 - e^{2\psi} A_{,z}^2 &=& 0 \, , \label{eqs2 m} \\ 
 2 \psi_{,z}^2 r^2 - e^{2\psi} A_{,r}^2 &=& 0 \, , \label{eqs3 m} \\ 
 \nabla^2 \psi - \nabla \psi \cdot \nabla \psi &=& 0 \, , \label{psi psi m} 
\end{eqnarray} 
where it can be seen, from the first tree equations, that it is possible to find relations between the electromagnetic four potential $A_\alpha$ and the metric function $\psi$ of the form
\begin{eqnarray}
A_{,r} =  \sqrt{2} e^{-\psi} r \psi_{,z} \, , \qquad
A_{,z} = - \sqrt{2} e^{-\psi} r \psi_{,r} \label{Ar_Az} \, ,
\end{eqnarray}
whose integrability conditions are given by equation (\ref{psi psi m}). From equation (\ref{eq: [F]}) we obtain a differential equation for $M_z$ 
\begin{eqnarray}
-\psi_{,r}M_z - {M_z}_{,r} + 2\sqrt{2} \psi_{,r}=0 \, ,
\end{eqnarray}
whose solution is given by $M_{z} = 2\sqrt{2} \left( 1 - e^{-\psi}  \right) $, where the constant of integration has been chosen equal to zero to guarantee a spacetime asymptotically flat. 

On the other hand, since we have chosen the metric (\ref{ds2}), accordingly equation (\ref{Tsup}) can be used to obtain the eigenvalues of the energy-momentum tensor in terms of the metric function $\psi$, that is
\begin{eqnarray}
& & S_{tt} = 4 e^{3\psi} \psi_{,z} \, , \quad   
 S_{rr} = -\sqrt{2} r e^{-\psi}\psi_{,z}M_z \, , \quad \\
& & S_{\phi \phi}= r^2 S_{rr} \, , \quad  
 S_{z z} = 0.
 \end{eqnarray}
However, their physical meaning can only be explored by the co-movil observer,  therefore, we calculate its components in the tetrad (\ref{tetrada}), through the relation $S_{(\alpha) (\beta)} = e^{\mu}_{(\alpha)} e^{\nu}_{(\beta)} S_{\mu \nu} $, which gives the following expressions for the surface energy density and pressures of the disk
\begin{eqnarray}
& & \sigma = S_{(t)(t)} = 4 e^{\psi} \psi_{,z}  \label{Dens_DM} \, ,  \\
& & p_r = S_{(r)(r)} =  - \sqrt{2}\ e^{\psi} \psi_{,z} M_z \, ,  \label{pr_DM}  \\ 
& & p_\phi = S_{(\phi)(\phi)} =  p_r \, , \label{pphi} \\ 
& & p_z = S_{(z)(z)} =  0 \label{pz_DM}  \, .
\end{eqnarray}
Additionally, we calculate the normal component of the magnetization vector, 
\begin{eqnarray}
M_{(z)} =  M_\beta e^\beta_{(z)}= M_z e^\psi = 2\sqrt{2} \left(  e^\psi - 1  \right) \, ,
\end{eqnarray}
and the components of the magnetic field through the relation
\begin{eqnarray}
B_{(\alpha)} = B_\mu e^\mu_{(\alpha)} = \frac{1}{2} \epsilon_{\mu\beta\lambda \nu} F^{\lambda \nu} u^\beta e^\mu_{(\alpha)} \, ,
\end{eqnarray}
which gives
\begin{eqnarray}
& & B_{(r)}=  - \sqrt{2}\ e^{\psi} \psi_{,r} = A_{,z} e^{2 \psi}/r \, , \label{B(psi)_r}  \\
& & B_{(z)}=  - \sqrt{2}\ e^{\psi} \psi_{,z} = - A_{,r} e^{2 \psi}/r \label{B(psi)_z} \, .
\end{eqnarray}
Furthermore, the expression for the total mass for an stationary and asymptotically flat spacetime \cite{RelativistikToolkit}, allows us to obtain the mass of the disk in terms of the eigenvalues
\begin{eqnarray}
M = 2 \pi \int_{0}^\infty \left( \sigma + p_r + p_{\phi} + p_z \right) r e^{-\psi} \ \mathrm{d} r   \, , \label{MDiscos}
\end{eqnarray}
in such a way that by substituting the expressions (\ref{Dens_DM}), (\ref{pr_DM}) and (\ref{pz_DM}) on this last equation, it can be written as
\begin{eqnarray}
M = 8 \pi \int_{0}^\infty \psi_{,z}\left( 2 e^{-\psi} -1 \right) r \mathrm{d}r   \, .  \label{MasaDM}
\end{eqnarray}

On the other hand, the equation (\ref{psi psi m}) becomes in $\nabla^2 \left( e^{-\psi} \right) = 0 $, which is the Laplace equation $\nabla^2 U = 0 $.  Therefore to guarantee the asymptotically flatness of the space time,  the metric function $\psi$ can be chosen as, $e^{-\psi} = 1 - U$, with $U=U(r,z)$ a solution of the Laplace equation. Accordingly, the physical quantities of the disk in terms of the function $U$ are
\begin{eqnarray}
& & \sigma  =  \frac{4 U_{,z}}{(1-U)^2} \, ,  \quad
 p_r = p_\phi = - \frac{4 U U_{,z}}{(1 - U)^2} \, , \label{ec:sigma_p(U)}  \\ 
& & M_{(z)}  =   \frac{ 2 \sqrt{2} U }{1 - U} \, ,  \quad
 B_{(r)} = -  \frac{\sqrt{2}U_{,r}}{(1 - U)^2}  \, , \label{ec:Mz_Br(U)} \\ 
& & B_{(z)} = -  \frac{\sqrt{2}U_{,z}}{(1 - U)^2}  \, , \quad
 M = 8 \pi \int_{r=0}^\infty \frac{{U}_{,z} \left(1 - 2U\right)r \mathrm{d}r }{\left( 1 - U \right)} \, . \qquad \label{ec:MasaDM(U)}
\end{eqnarray}

The most general axially symmetric solution in cylindrical coordinates of the Laplace equation is
\begin{eqnarray}
U_{n} = -\sum^n_{l=0} \frac{C_l P_l\left( z/R \right)}{R^{l+1}}  \, ,  \label{U(PL)}
\end{eqnarray}
with $C_l$ constants, $P_l(\cos \theta)$ the Legendre polynomials and $ R = \sqrt{r^2+z^2} $. This solution, which guarantees that the spacetime is asymptotically flat, is continuous with continuous derivatives.
In order to obtain the razor-thin disk, we need to introduce a discontinuity on its first derivative, which will be transported into the metric and the electromagnetic tensor. To do so, we apply the ``displace, cut and reflect'' method, which has been used in \cite{Kuzmin,Toomre} to generate disk-like sources in the frame of Newtonian gravity. This method is based on the transformation $ z \rightarrow |z| + a$, which is applied to the function $U_n$, whit $a$ a positive constant. Accordingly, we now have
\begin{eqnarray}
U_{n} = -\sum^n_{l=0} \frac{C_l P_l \left((|z| + a)/R \right)}{R^{l+1}}  \, ,  \label{U(PL)transf}
\end{eqnarray}
with $R=\sqrt{r^2+(|z| + a)^2}$. Hence, the surface energy density, pressure, magnetic field components and $z$-magnetization can be written in terms of (\ref{U(PL)transf}), see appendix \ref{apendice1}. It is worth mentioning that electrically polarized solutions were presented in \cite{proceeding_electrico}. 

\section{Two particular solutions of magnetized razor-thin disks}
\label{Sec:Results}

Let us consider the most simple terms obtained by expanding the solution (\ref{U(PL)transf}) until $n=0$ and $n=1$, which are given respectively by
\begin{eqnarray}
& & U_{0}= -\frac{C_0}{R}  \, ,  \label{Ucero} \\
& & U_1= -\frac{C_{0}}{R} -\frac{C_{1}(|z| + a)}{R^3} \label{U1} \, ,
\end{eqnarray}
with $R=\sqrt{r^2+(|z| + a)^2}$, in such a way that the physical features of the disks are now in terms of the coefficients $C_i$ (Please note that $C_0$ in equations (\ref{Ucero}) and (\ref{U1}) can be different.) and the constant $a$. For the density, pressure and magnetization vector the expressions are evaluated in $z=0$ since they are zero elsewhere.  The values of the coefficients need to be selected to satisfy the energy conditions \cite{RelativistikToolkit,OscarEnergy,OscarEnergy2}, 
\begin{eqnarray}
\sigma \geq 0 \, ,  \label{Condicion1} \\
\sigma + p_i \geq 0  \, , \label{Condicion2}  \\ 
|\sigma| \geq | p_i |   \, , \label{Condicion3} \\ 
\sigma + p_r + p_{\phi} + p_{z} \geq 0 \, . \label{Condicion4}
\end{eqnarray}
which guarantee a physically acceptable (standard) behavior of the matter and its gravitational field. Here (\ref{Condicion1}) and (\ref{Condicion2}) correspond to the weak energy condition, which states that the density of energy of any matter distribution must be nonnegative. Equation (\ref{Condicion2}) and (\ref{Condicion4}) stand for the strong energy condition  which ensures the attractive nature of the gravitational field. Finally, (\ref{Condicion1}) and (\ref{Condicion3}) are the dominant energy condition, which guarantees that the energy flux density vector must be a future oriented time-like or null vector.  

In the case of the first model $(n=0)$, $C_0 = a$ is the only acceptable value of the parameter and in the second one $(n=1)$, we have found that they need to satisfy the following inequalities
\begin{eqnarray}
C_1 \leq   0  \, ,  \quad  
C_0 \geq  2|C_1|  \, ,  \quad    
R_0^3  \geq   \left( C_0 R_0^2 + C_1 \right) \, , \label{relacionesEnergyConditions}
\end{eqnarray}
where $R_0 = \sqrt{r^2 + a^2}$. Since the only free parameter is the constant $a$, it can be used to adimensionalise the solution through the following relations
\begin{eqnarray}
\hspace{-0.5cm} r = a \bar{r} \, ,    \quad  
C_{0} = a \bar{C_{0}} \, ,   \quad C_1 = a^2 \bar{C_1} \, , \quad
R_0 = a \bar{R_0}  \, , \label{adim_coef}
\end{eqnarray}
and in terms of them, the surface energy density, pressures, magnetization, magnetic fields and the mass are presented in appendices (\ref{apendice_n0}) and (\ref{apendice_n1}). 

\subsection{Density, pressure and magnetization profiles\label{subsec:PerfilesDensidad}}

\begin{figure}
\centering
\includegraphics[width = 0.6\textwidth]{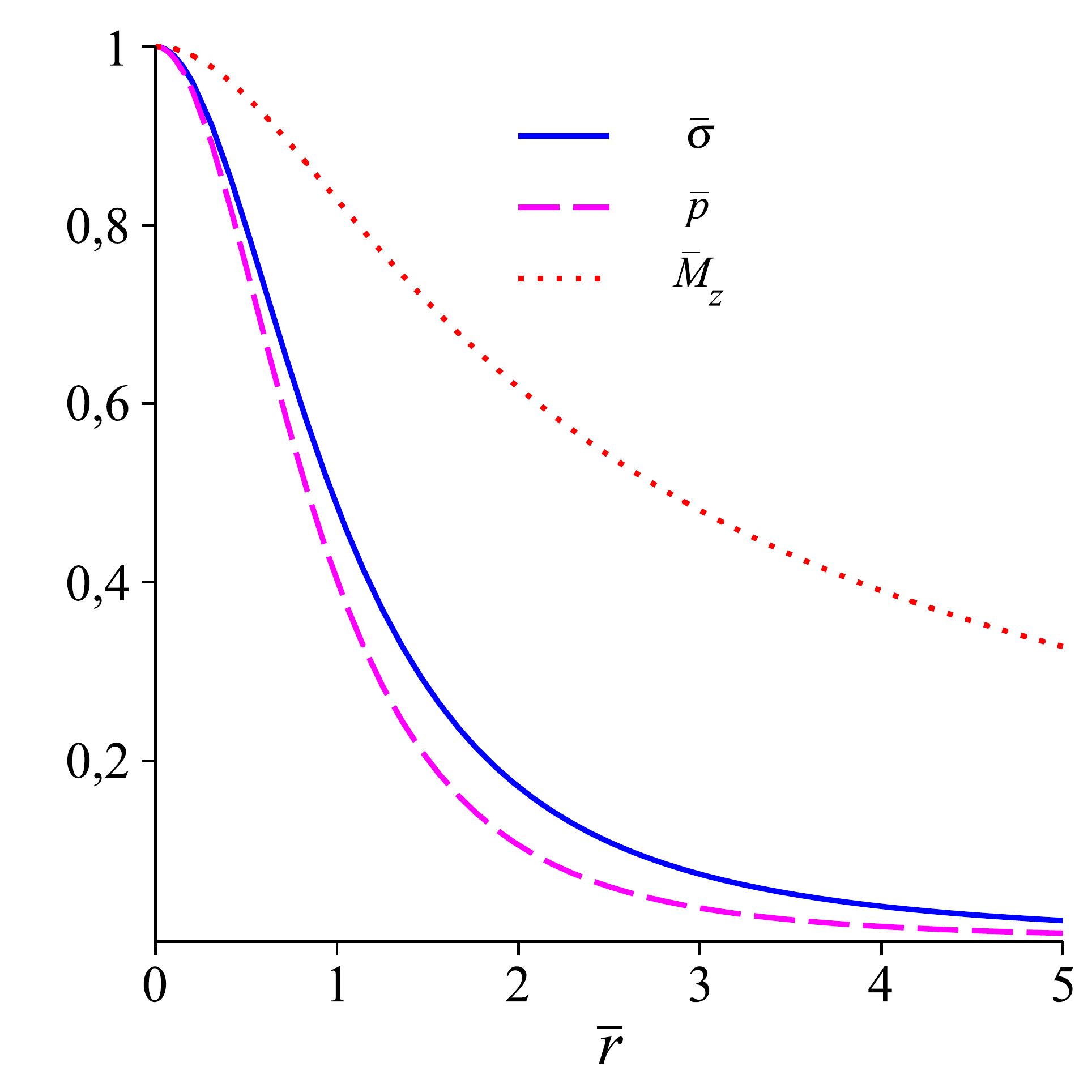}
\caption{Surface energy density $\bar{\sigma}$, pressure $\bar{p}$ and $z$-magnetization $\bar{M}_z$ of the model $n=0$, using $C_0 =1$. } \label{fig_Densidad_presion_magnetizacion_U0}
\end{figure}

In order to analyze the physical behavior that the solutions exhibit, we plot in figure \ref{fig_Densidad_presion_magnetizacion_U0} the surface energy density $\bar{\sigma}$, radial and azimuthal pressures $\bar{p_r}$ = $\bar{p_\phi} = \bar{p}$ and the $z$-component of the magnetization $M_{(z)}$ for the model $n=0$ taking $\bar{C_0} = 1$ (since it is the only value that makes the energy density and pressures to be in agreement with the energy conditions). The plot shows that these quantities are everywhere positive, regular, and have their maximum value at the center of the disk. On the other hand, as $r$ increases, they tend to zero, as should be expected. In addition, it can be noticed that the magnetization does not decrease in the same proportion than the density and the pressure. For instance, for a radius $r\approx 5$, it has decreased a $67 \%$, while the density and pressure diminished their values in a $97.9\%$ and $99.3\%$ respectively. This last result may be in agreement with the measurements of significant magnetic fields at the intergalactic space 
\cite{2002PhT....55l..40K}. 

\begin{figure}
\centering
\includegraphics[width = 0.6\textwidth]{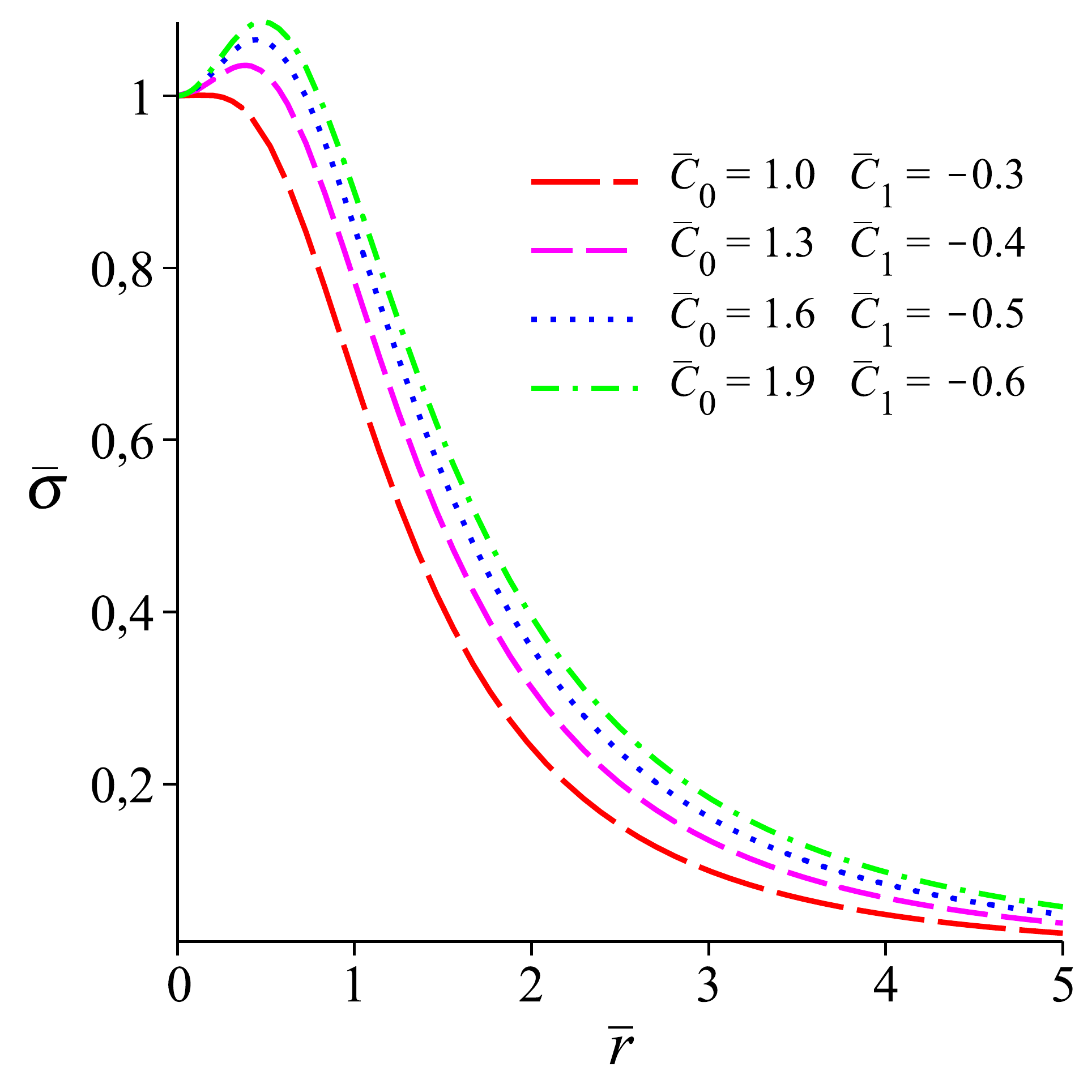}
\includegraphics[width = 0.6\textwidth]{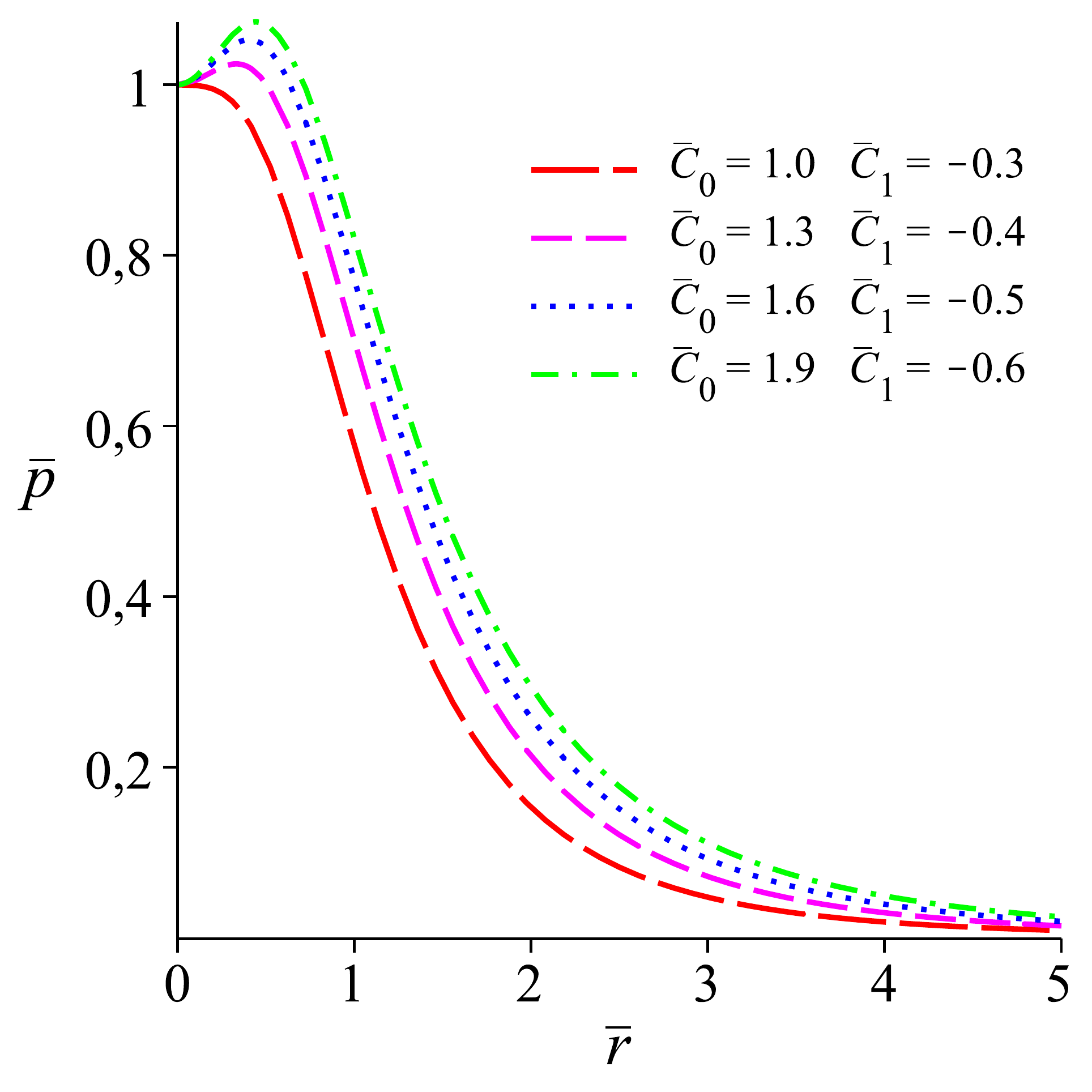}
\includegraphics[width = 0.6\textwidth]{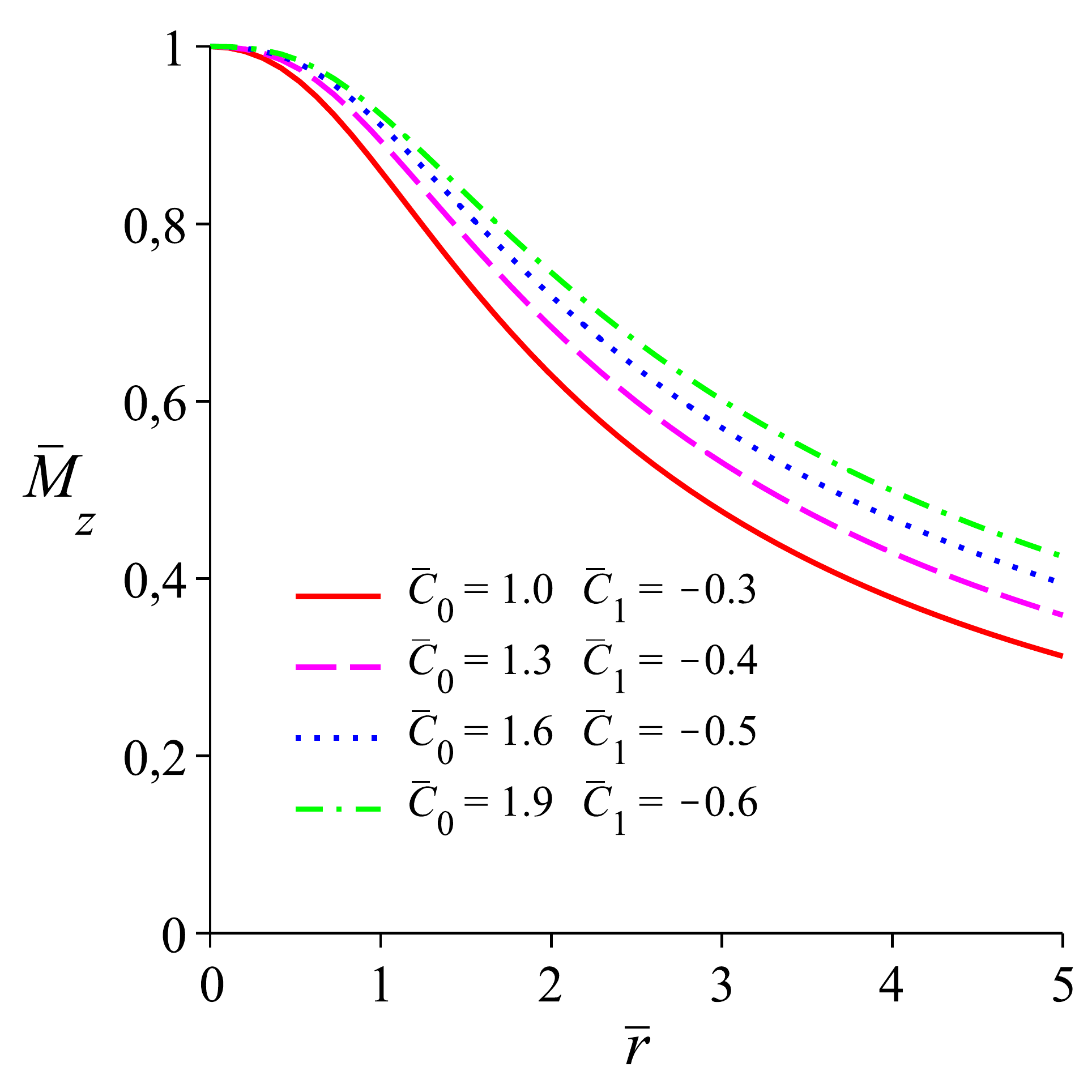}
\caption{Surface energy density $\bar{\sigma}$, pressure $\bar{p}$ and $z$-magnetization $\bar{M}_z$ of the model $n=1$.} \label{fig_Densidad_presion_magnetizacion_U1}
\end{figure}

Concerning to the second model, $n=1$, there exist several combinations of coefficients $\bar{C}_0$ and $\bar{C}_1$ according to the relations (\ref{relacionesEnergyConditions}). We chose the following [$\bar{C}_0 = 1.0 $, $\bar{C}_1 = -0.3 $], [$\bar{C}_0 = 1.3 $, $\bar{C}_1 = -0.4 $], [$\bar{C}_0 = 1.6 $, $\bar{C}_1 = -0.5 $] and [$\bar{C}_0 = 1.9 $, $\bar{C}_1 = -0.6 $]. With those values, we plot in figure \ref{fig_Densidad_presion_magnetizacion_U1}  the surface energy density $\bar{\sigma}$, pressures $\bar{p_r}$ = $\bar{p_\phi} = \bar{p}$ and $z$-magnetization $\bar{M}_{(z)}$, respectively.  In this case, we have also found that the $z$-magnetization does not drop that dramatically as the density and pressure, and it also has a maximum value in the center of the disk. However, in this model we can see that the profiles of density and pressure display a different behavior: for increasing values of $\bar{C}_0$, the radius at which the maximum of the curve is located, is slightly displaced from the center, while it maximum value also increases. This negative gradients could origin different dynamics within the center of the configuration and may be interesting to study through numerical simulations.     

\subsection{Magnetic field lines\label{Subsec:Lineas}}

\begin{figure}
\centering
\includegraphics[width=0.6\textwidth]{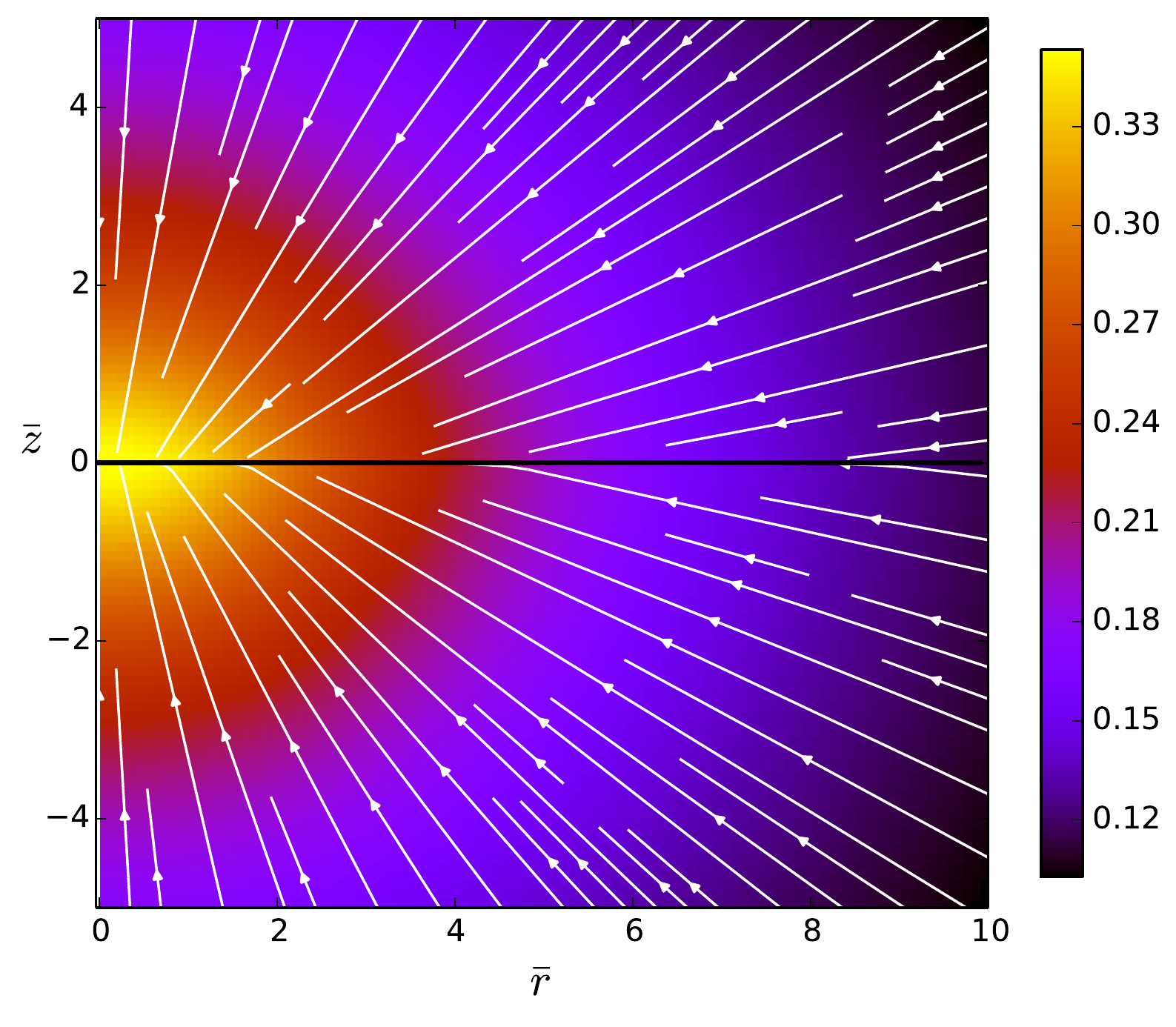}
\caption{Magnetic field lines for the model $n=0$, using $\bar{C_0}=1$.} \label{fig_lines_n0}
\end{figure}

\begin{figure*}
\centering
\includegraphics[width=0.45\textwidth]{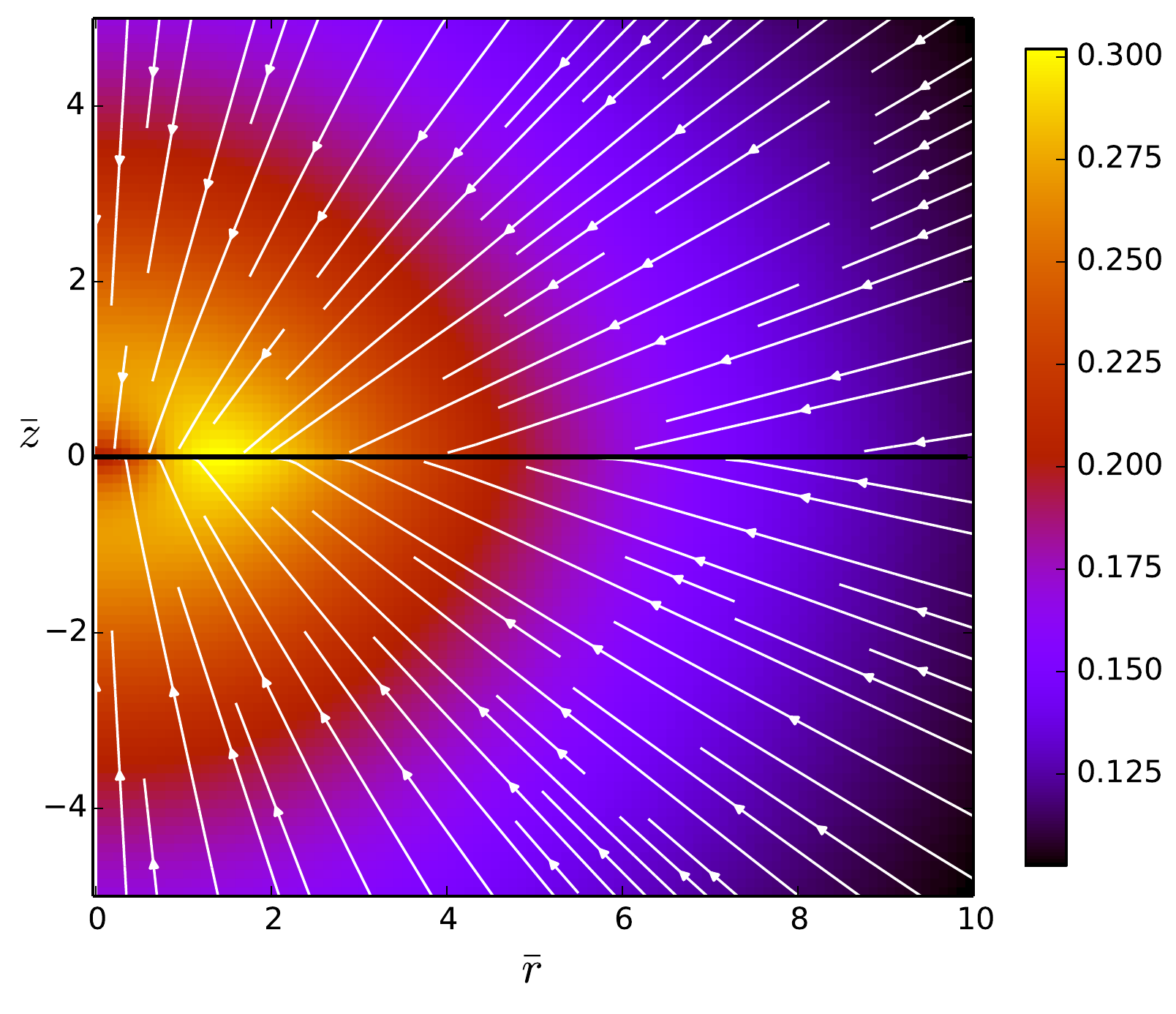}
\includegraphics[width=0.45\textwidth]{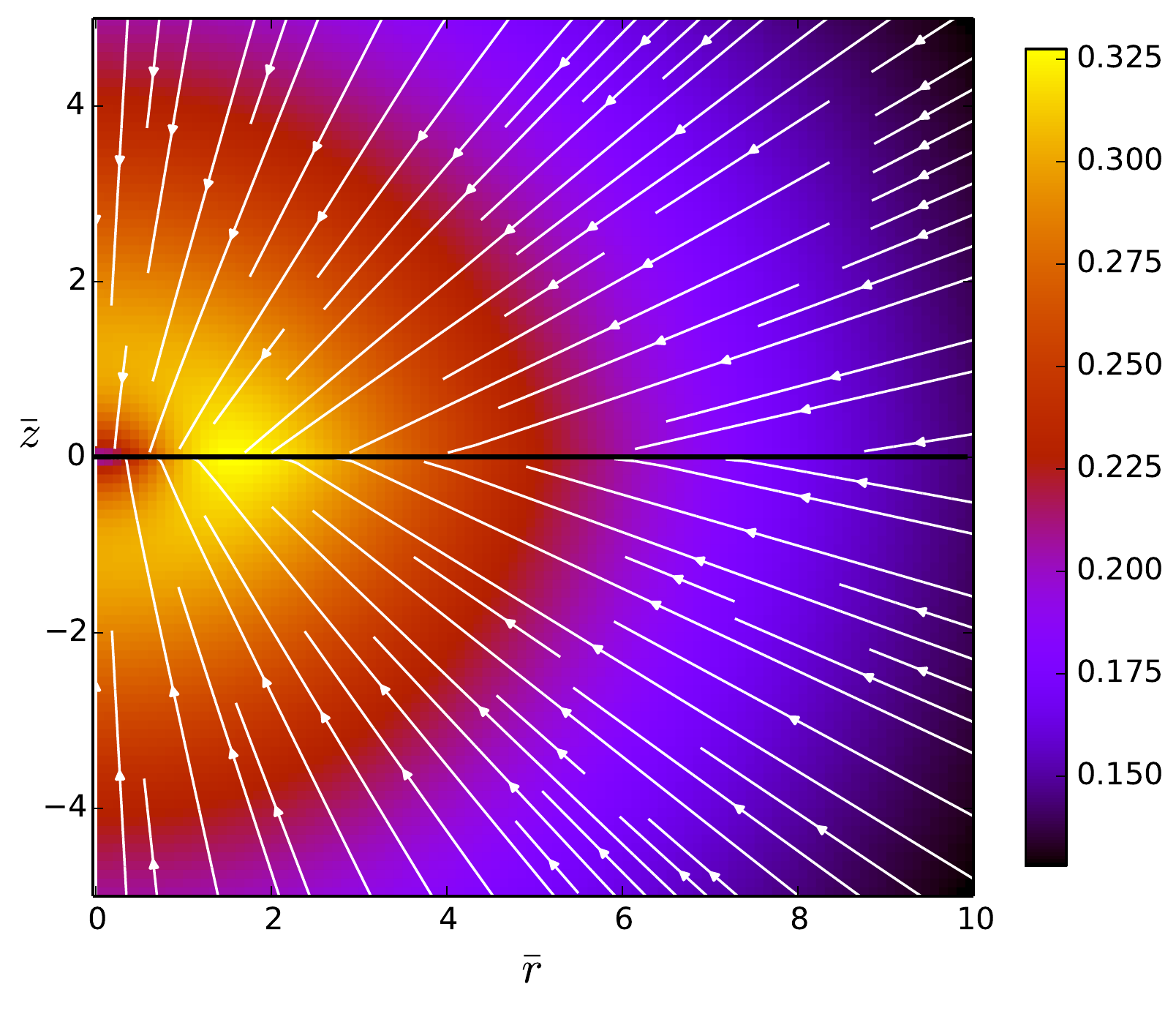}
\includegraphics[width=0.45\textwidth]{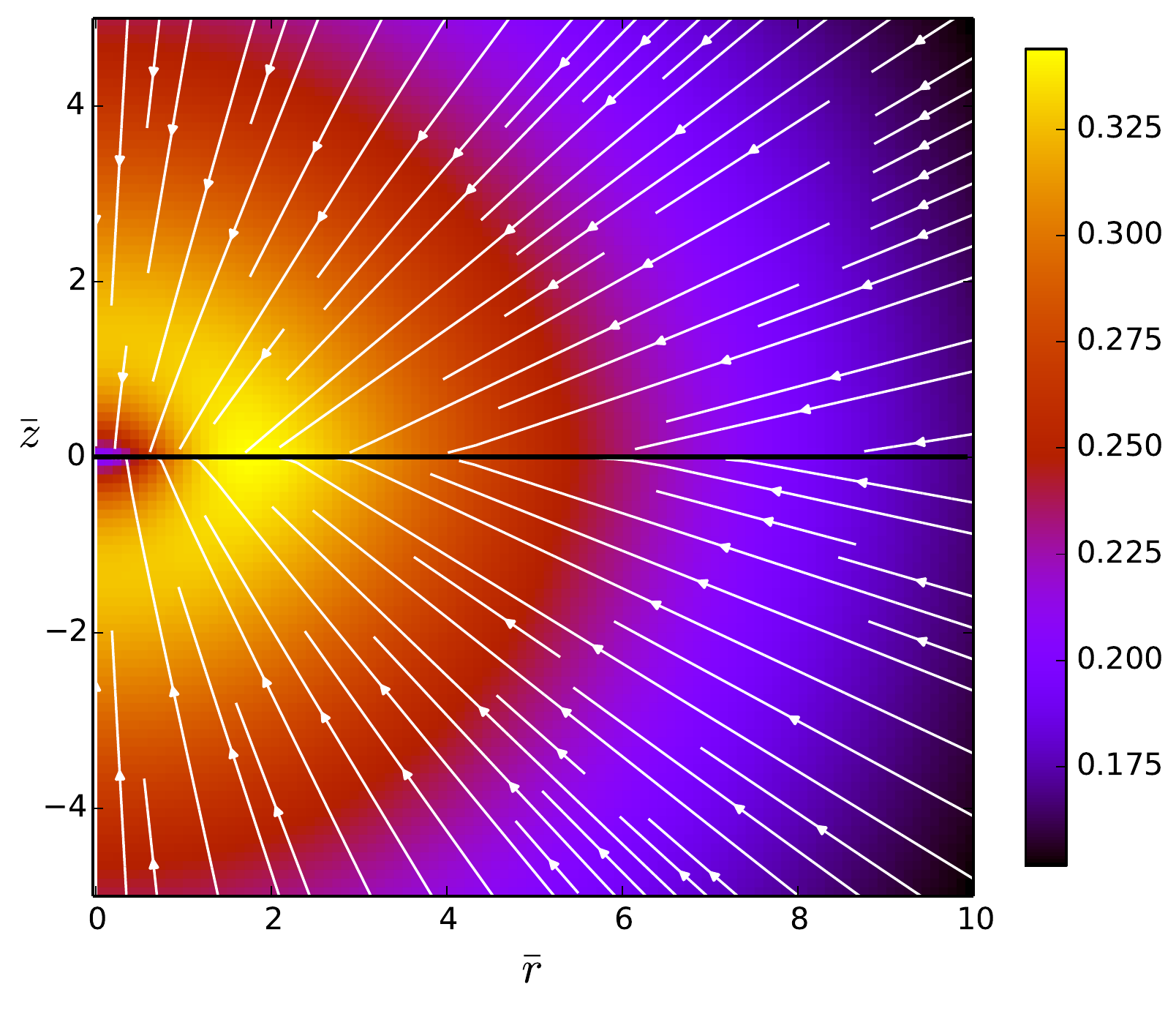}
\includegraphics[width=0.45\textwidth]{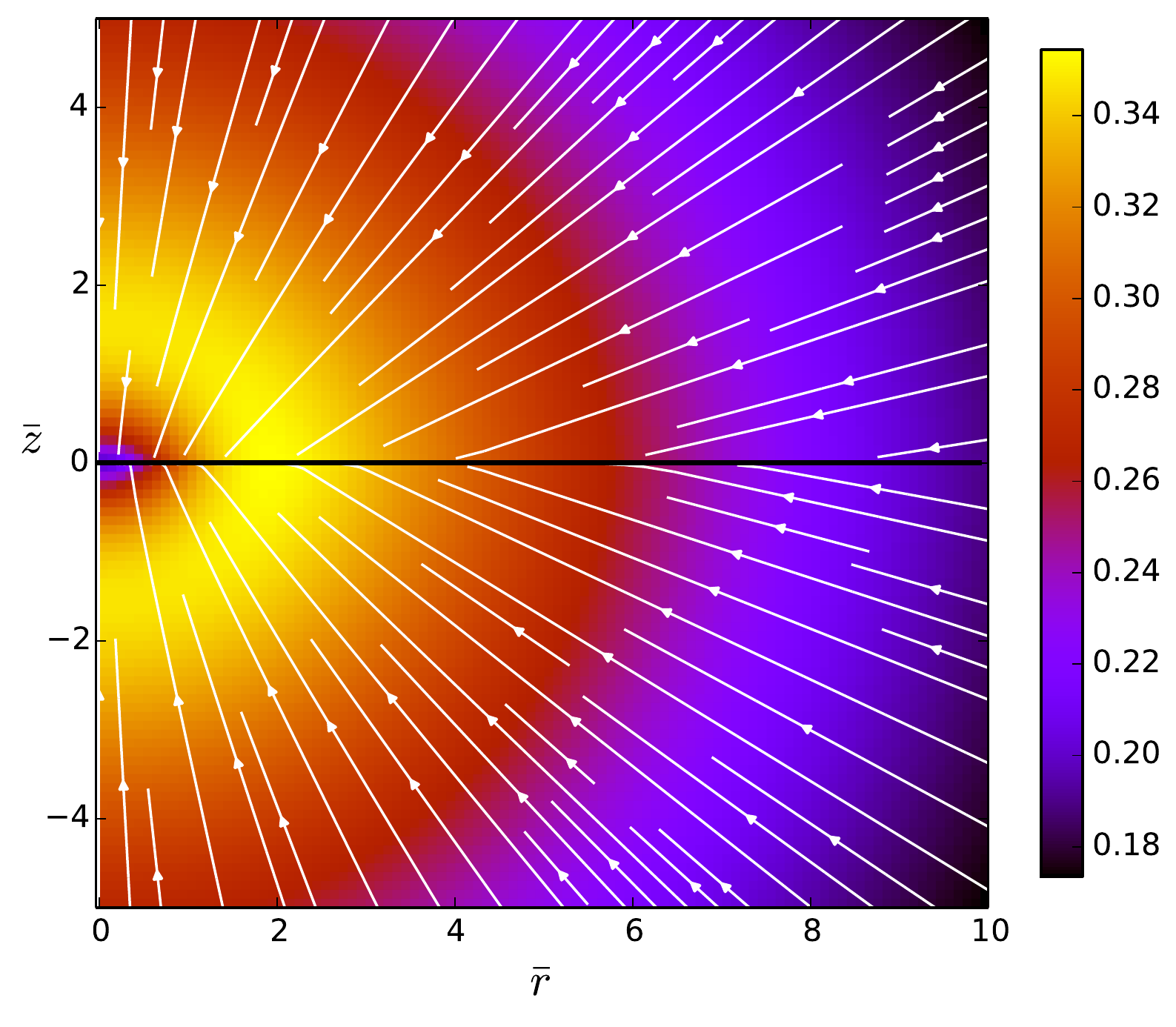}
\caption{ \label{fig_lines_n1} Magnetic field lines for the model $n=1$, for the following combinations of the parameters $\bar{C}_0 = 1.0 $, $\bar{C}_1 = -0.3 $ (upper-left panel), $\bar{C}_0 = 1.3 $, $\bar{C}_1 = -0.4 $ (upper-right panel), $\bar{C}_0 = 1.6 $, $\bar{C}_1 = -0.5 $ (bottom-left panel) and $\bar{C}_0 = 1.9 $, $\bar{C}_1 = -0.6 $ (bottom-right panel). } 
\end{figure*}

From the expressions of the magnetic field components given in  appendices \ref{apendice_n0} and \ref{apendice_n1} for the models $n=0$ and $n=1$ respectively, we plot the magnetic field lines along with a colormap of the magnitude of the magnetic field. In figure \ref{fig_lines_n0} they are presented for the fist model, where there it can be seen that the lines follow straight paths and the magnetic field is more intense in the center. As it is expected from the procedure used to obtain the solutions, they have symmetry of reflection through the plane of the disk and also are axially symmetric. For the second model, we also plot in figure \ref{fig_lines_n1} the magnetic field lines and the colormap of the magnetic field magnitude,  using the same combinations of parameters than for the density, pressure and $z$-magnetization, which are [$\bar{C}_0 = 1.0 $, $\bar{C}_1 = -0.3 $], [$\bar{C}_0 = 1.3 $, $\bar{C}_1 = -0.4 $], [$\bar{C}_0 = 1.6 $, $\bar{C}_1 = -0.5 $] and [$\bar{C}_0 = 1.9 $, $\bar{C}_1 = -0.6 $]. In this cases the lines behave in a similar way, they are straight and have a similar inclination. However, the intensity of the magnetic field is displaced from the center depending on the election of the parameters, this result is in agreement with the shift shown in the density and pressures shown before. Since the lines do not deflect in the outer radius, they could be qualitatively in agreement with the observed magnetic fields in the intergalactic space, and also with the low rate of decaying of the $z$-magnetization of the models. 

\subsection{Rotation Curves\label{Subsec:Rotacion}}
\begin{figure}
\centering
\includegraphics[width = 0.45\textwidth]{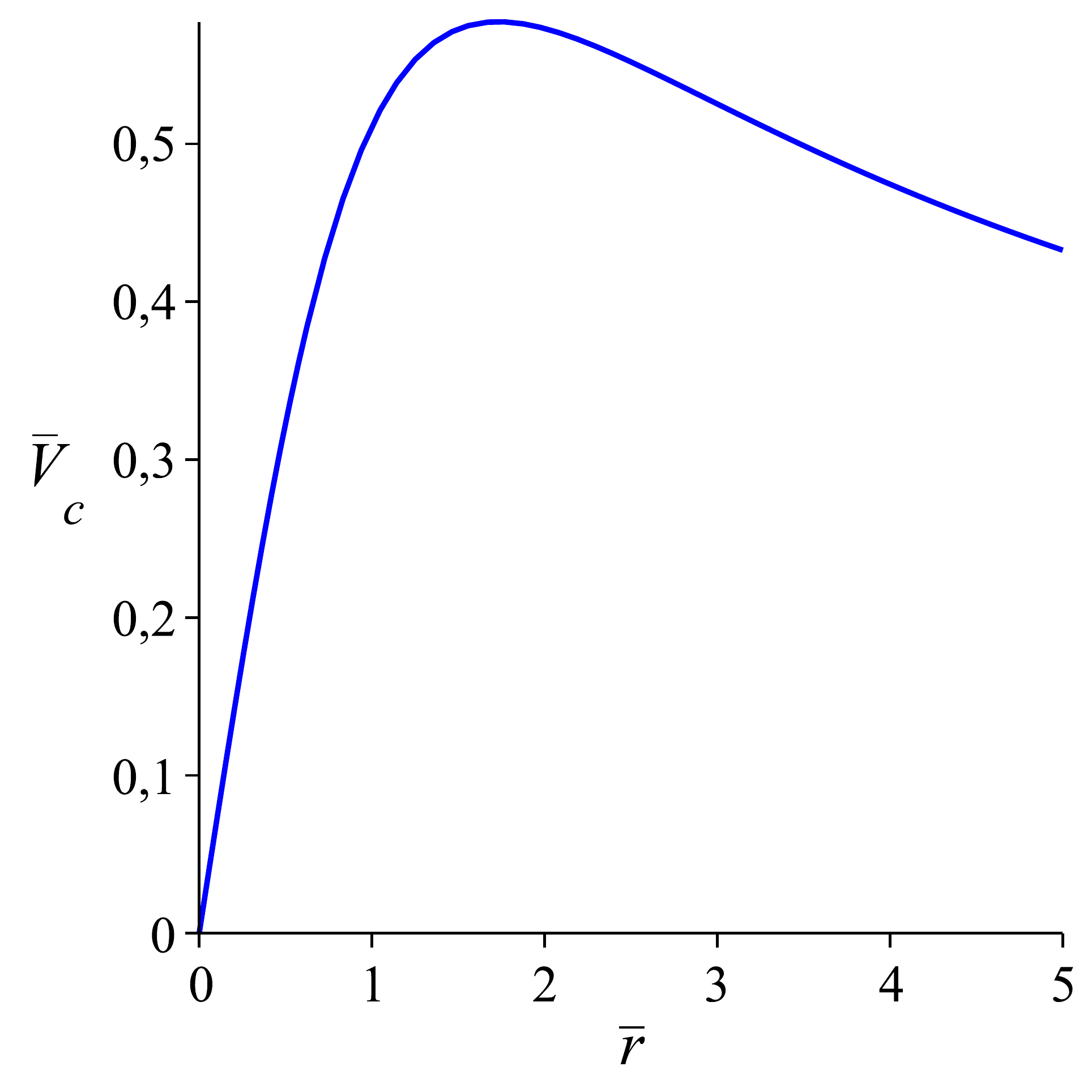}
\includegraphics[width = 0.45\textwidth]{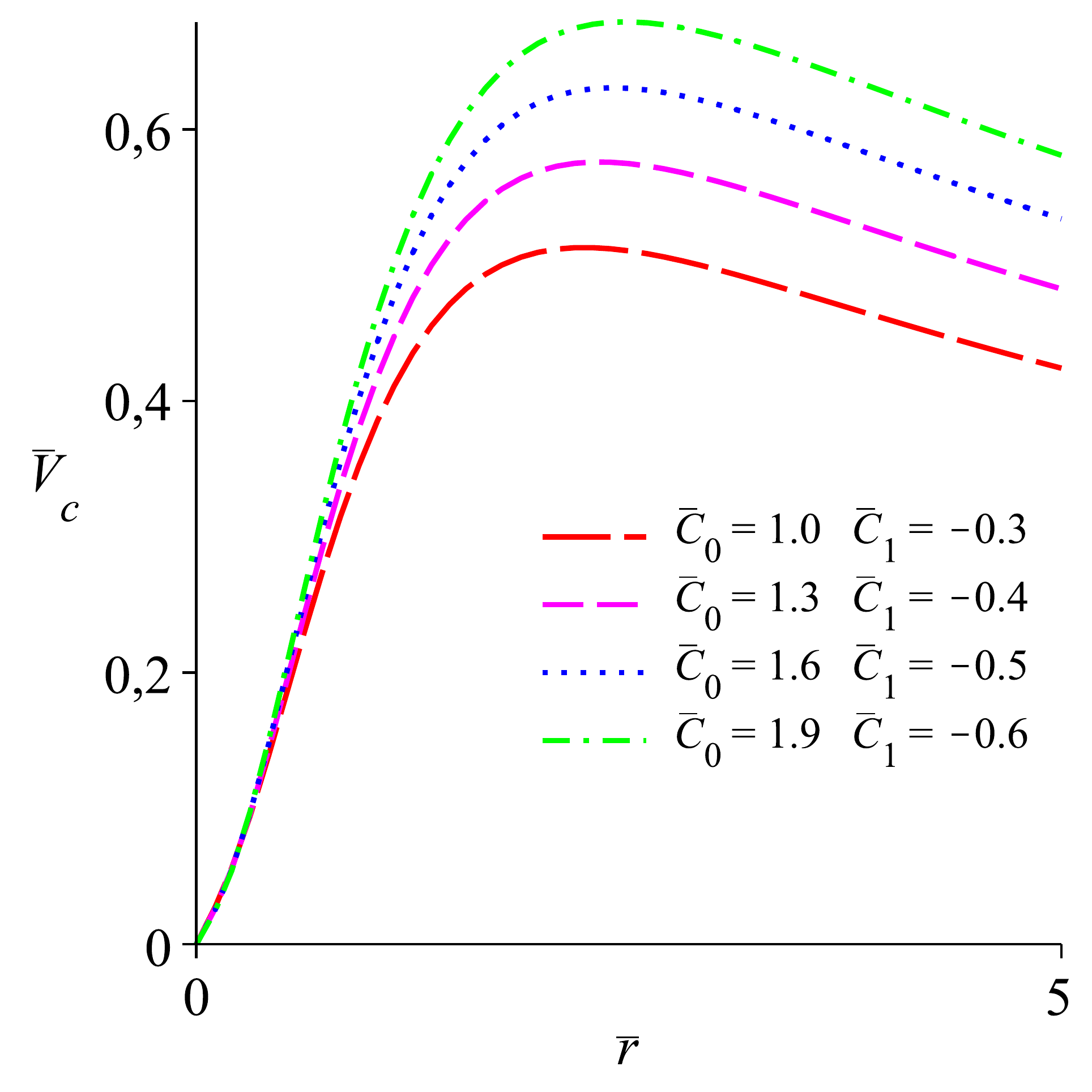}
\caption{Rotation curve of the models $n=0$ and $n=1$, respectively. } \label{fig_vel_rot}
\end{figure}

 An important aspect that can be studied in this system are the circular geodesics along the plane of the disk (the $z=0$ surface) in order to determine qualitatively its resemblance with observed rotation curves. In  appendix \ref{apendice_vc} we obtain the general expression for the rotational velocity for the models $n=0$ and $n=1$. Their plot is presented in figure \ref{fig_vel_rot} using the same radial range then the ones used in the plots of the other physical quantities, this is until $\bar{r}\approx 5$, since for this value most of the mass is concentrated. The displayed behavior of the curves is in agreement with the one predicted for rotation curves from Newtonian gravity for spiral galaxies \cite{Andromeda_Galaxy}, since they tend to be plane when $r$ increases.  
 
 \subsection{Convergence of the mass\label{Subsec:Masa}}
The convergence of the mass for every model of the family is proved in appendix \ref{apendice_mass}. This result is obtained by calculating the mass of the first and most simple model ($n=0$), which we found that is finite. Now, in regards of the ``Limit Convergence Test'' \cite{Apostol} the convergence of any model of the expansion is also proved.  
This has a great importance, since it demonstrates that the mass has a finite value no matter which value of $n$ is chosen, and validates the use of the models for galactic disks. 

\section{Conclusion}
\label{sec:conclusions}

An infinite family of exact solutions of the Einstein-Maxwell equations was obtained for a conformastatic spacetime, static and axially symmetric. The solutions describe thin magnetically polarized disks, and were obtained by solving the Einstein-Maxwell Field Equations for continuum media without the test field approximation, and assuming that the sources are razor thin disk of magnetically polarized matter. We analyzed the first two models, $n=0$ and $n=1$, and found they behave in a physically expected way, they satisfy the energy conditions and the energy density, radial pressure, azimuthal pressure and $z$-magnetization are everywhere positive, smooth, decay when the distance from the center increases. The $z$-magnetization decays more slowly compared with the density and pressures. Despite of the infinite extension of the disk, the value of the total mass converges and the physical features are concentrated in the central region as we could expect from a physical system. The rotation curves for circular orbits exhibited a typical behavior from a theoretical point of view.  This article serves as a starting point in the study of magnetically polarized matter within the exact solutions of the Einstein-Maxwell equations. Also, the solutions may be used to model simplistic astrophysical galactic disks, and may be used as initial data in relativistic numerical simulations.

\begin{acknowledgements}
A. N wants to thanks the financial support from COLCIENCIAS, Colombia, under the program ``Becas Doctorados Nacionales 647'' and Universidad Industrial de Santander.  F.D.L-C gratefully acknowledges the financial support from VIE-UIS, grant number 1822.  G.A.G. was supported in part by VIE-UIS, under Grants No.  1347 and No.  1838, and by COLCIENCIAS, Colombia, under Grant No. 8840.
\end{acknowledgements}



%
%

\appendix

\section{Analytical Expressions for the Physical Features of the disk of the infinite family}\label{apendice1}

When the solution of the Laplace equation (\ref{U(PL)transf}) is introduced in the equations (\ref{ec:sigma_p(U)}) to (\ref{ec:MasaDM(U)}), the general expressions for the energy density, radial and azimuthal pressures, magnetic fields components, magnetization and the mass are given by
with $R_0 = \sqrt{r^2 + a^2}$, $\xi_1$, $\xi_2$, $\xi_3$, $\xi_4$ and $\xi_5$ defined as
\begin{eqnarray}
& &\xi_1 = C_l(l+1) P_{l+1}(a/R_0)/R_0^{(l+2)} \, , \label{xi_1} \\
& &\xi_2 =  C_{m}  P_{m+1}^{'}(a/R_0)/R_0^{(m+3)}  \, ,\label{xi_2}  \\ 
& &\xi_3 =   C_l(l+1) P_{l+1}(a/R_0) / R_0^{(l+2)}  \, ,\label{xi_3} \\ 
& &\xi_4 = C_l P_l(a/R_0)/R_0^{l+1} \, , \label{xi_4}  \\ 
& &\xi_5 = C_l  P_l^{'}(a/R_0)/R_0^{(l+3)} \, . \label{xi_5} 
\end{eqnarray}


\section{Expressions for the model $n=0$ }\label{apendice_n0}

For the model $n=0$, the expressions for the energy density, radial and azimuthal pressures, magnetic fields components, magnetization and the mass are given by
\begin{eqnarray}
 & &\bar{\sigma}  = \frac{\sigma}{\sigma_0} = \frac{(1 + \bar{C}_0)^2 }{\bar{R_0} (\bar{R_0} + \bar{C}_0)^2}  \label{DMsigmanormU0}   \, , \\  
& & \bar{p_r} = \bar{p_{\phi}}= \frac{p_r}{p_{r_0}}= \frac{(1 + \bar{C}_0)^2 }{\bar{R_0}^2 (\bar{R_0} + \bar{C}_0)^2}  \, ,   \label{DMprnormU0}       \\  
 & & \bar{M}_{(z)} = \frac{M_{(z)}}{{M_{(z)}}_0} = \frac{1 + \bar{C_0}}{\bar{R_0} + \bar{C_0}} \, , \label{DMmagnetizacionznormU0}
\end{eqnarray}
with 
\begin{eqnarray}
\hspace{-0.5cm} \sigma_{0} = \sigma|_{(r=0)} \, ,  \    
p_{r_0} =  p_r|_{(r=0)}  \, ,  \
{M_{(z)}}_0 = \left. M_{(z)}\right|_{r=0}. \label{Mz0}
\end{eqnarray}
It is worth mentioning that this expressions are evaluated in $z=0$ since they are zero elsewhere. 


\section{Expressions for the model $n=1$ }\label{apendice_n1}

For the model $n=1$, the expressions for the energy density, radial and azimuthal pressures, magnetic fields components, magnetization and the mass are given by
\begin{eqnarray}
 & & \bar{\sigma} = \frac{\sigma}{\sigma_0} =  \frac{\bar{R_0}  \Delta_8  \left( \Delta_2  \right)^2 }{\left( \Delta_3 \right)^2 \Delta_9} \label{DMsigmanormU1} \, ,  \\  
 & & \bar{p_r}  =  \bar{p_{\phi}}  =  \frac{p_r}{p_{r_0} }  =  
 \frac{ \Delta_{10}    \Delta_8  (\Delta_2)^2  }{\bar{R_0}^2 ( \Delta_3  )^2  \Delta_9  \left( \Delta_{11} \right) }    \, , \label{DMpresionrnormU1} \\  
& & \bar{M}_{z} = \frac{M_z}{M_{z_0}}= \frac{ \Delta_{10}  \Delta_2 }{  \Delta_3  \Delta_{11} }     \, , \label{DMmagnetizacionnormU1}  
\end{eqnarray}
where
 \begin{eqnarray}
& & \Delta_1 = \bar{C_0} \bar{R_0}^4   +  2\bar{C_1}^2  \bar{R_0}  +  \bar{C_1}  \left[ (\bar{R_0}^2  +  1) -  \bar{r}^4 \right]  \, , \label{Delta_1} \\ 
& &\Delta_2 =  1  +  \bar{C_0}  +  \bar{C_1}  \, , \label{Delta_2} \qquad \qquad
\Delta_3 = \bar{R_0}^3  +  \bar{C_0} \bar{R_0}^2   +  \bar{C_1} \, ,  \qquad \label{Delta_3}\\
& & \Delta_4 = \bar{C_0}  +  2 \bar{C_1}(1  +   \bar{C_1} ) \, , \label{Delta_4} \hspace{0.8cm}
\Delta_5 = \bar{C_0} \bar{R_0}^2 + 3\bar{C_1} \, ,  \label{Delta_5} \\
& & \Delta_6 = \bar{C_0} \bar{R_0}^2 - \bar{r}^2 \bar{C_1} \, , \label{Delta_6} \hspace{1.5cm}
\Delta_7 = \bar{C_0} + 3\bar{C_1} \, ,\label{Delta_7}\\
& & \Delta_8 = \bar{C_0} \bar{R_0}^2 + \bar{C_1}\left(2 - \bar{r}^2\right)  \, , \hspace{0.3cm}  
\Delta_9 = \bar{C_0} + 2 \bar{C_1} \, , \\  
& & \Delta_{10} = \bar{C_0} \bar{R_0}^2 + \bar{C_1} \, , \hspace{1.7cm}  
\Delta_{11} = \bar{C_0} + \bar{C_1} \, ,
\end{eqnarray}
with all the expressions evaluated in $z=0$ since they are zero elsewhere.

\section{Convergence of the mass }\label{apendice_mass}

The general expression for the mass of the disk is given by (\ref{ec:MasaDM(U)}), with it we calculate the total mass for the model $n=0$, then we have
\begin{eqnarray}
M = 8 \pi  \left(  2 a \ln \left( \frac{C_0+a}{a} \right) - C_0 \right) \, ,
\end{eqnarray}
from where we can see that their mass have finite values.  On the other hand, the general integral factor of the mass of the disk is given by  
\begin{eqnarray}
f_n = \frac{  \left[ \sum\limits_{l=0}^n \xi_1  \right] \left[1 + 2 \sum\limits_{l=0}^n \xi_4 \right] }{\left[ 1 + \sum\limits_{l=0}^n  	\xi_4 \right]} \, ,
\end{eqnarray}

and we can easily prove that 
\begin{eqnarray}
\lim_{R \rightarrow \infty} \frac{f_{n+1}}{f_n} = 1 \, . \label{limit=1}
\end{eqnarray}
Then, in regard of the `` Limit Convergence Test'' \cite{Apostol}, the convergence of the mass for the first model and the limit (\ref{limit=1}) prove that the value of the masses of the disks for all the infinite models have a finite value. 
 
\section{Rotational velocity}\label{apendice_vc}

The general velocity of a proof mass in a circular orbit is given by $u^\alpha = u^t \left( 1 , \Phi , 0 , 0 \right)$, 
where $\Phi = u^\phi/u^t $ is the angular velocity. By imposing the normalization condition $g_{\alpha \beta} u^\alpha u^\beta = 1 $ we obtain the relation
\begin{eqnarray}
\left( u^t \right)^2 = \frac{e^{-2\psi}}{1 - {V_c}^2 } \, ,
\end{eqnarray}
where
\begin{eqnarray}
{V_c} = e^{-2\psi}r\Phi \, , \label{V_c}
\end{eqnarray}
is the rotational velocity. On the other hand, the geodesic equation
\begin{eqnarray}
\frac{d u_\alpha}{d\tau} = \frac{1}{2} g_{\mu \nu , \alpha} u^\mu u^\nu \, ,
\end{eqnarray}
solved for $\alpha = r$ implies
\begin{eqnarray}
\frac{d u_r}{d\tau} = \frac{1}{2} \left[ {g_{t t}}_{,r} (u^t)^2 + {g_{\phi \phi}}_{,r} (u^\phi)^2 \right] = 0 \, .
\end{eqnarray}
From the last expression we solve $\Phi$,
\begin{eqnarray}
\Phi = \frac{\psi_{,r} \ e^{4\psi}}{r ( 1 - r \psi_{,r}  ) } \, , 
\end{eqnarray}
and replace it on equation (\ref{V_c}) to obtain a general expression for the rotational velocity,
\begin{eqnarray}
V_{c} = \sqrt{\frac{ r \psi_{,r} }{ 1 - r \psi_{,r} }   } \, .
\end{eqnarray}
Now, in order to regard its behavior in the models $n=0$ and $n=1$ we substitute the metric function by using $e^{-\psi} = 1 - U$, this is
\begin{eqnarray}
{V_n}_{c} = \sqrt{\frac{ r {U_n}_{,r} }{ 1 - U_n - r {U_n}_{,r} }   } \, .
\end{eqnarray}
Accordingly, for the model $n=0$ the rotational velocity is 
\begin{eqnarray}
{V_0}_c = \bar{r} \sqrt{\frac{ \bar{C_0} }{ \bar{R_0}^3 + \bar{C_0} } } \, ,
\end{eqnarray}
and for the model $n=1$ the rotational velocity is  
\begin{eqnarray}
{V_1}_c = \bar{r} \sqrt{ \frac{ \bar{C_0} \bar{R_0}^2 + 3 \bar{C_1}  }{ \bar{R_0}^5 + \bar{C_0} \bar{R_0}^2 - \bar{C_1} ( 2\bar{r}^2 - 1 )  } } \, .
\end{eqnarray} 

\end{document}